\documentclass[11pt,a4paper]{article}
\usepackage{jheppub}

\usepackage{amsmath}
\usepackage{amssymb}
\usepackage{mathrsfs}
\usepackage{graphicx}
\usepackage{dcolumn}
\usepackage{upgreek}
\usepackage{color}
\usepackage{xcolor}
\usepackage{url}
\usepackage[toc,page]{appendix}
\usepackage[normalem]{ulem}
\usepackage{siunitx}

\usepackage{pdfpages}

\graphicspath{{figs/}}

\title{Can we discover lepton number violation with LHC far detectors?}

\author[\,a]{Ying-nan Mao}
\emailAdd{ynmao@whut.edu.cn}
\affiliation[a]{Department of Physics, School of Science,
Wuhan University of Technology, 430070 Wuhan, Hubei, China}

\author[\,a]{, Kechen Wang}
\emailAdd{kechen.wang@whut.edu.cn}

\author[\,b,\,c,\,1]{and Zeren Simon Wang\note{Corresponding author.}}
\emailAdd{wzs@mx.nthu.edu.tw}
\affiliation[b]{Department of Physics, National Tsing Hua University, Hsinchu 300, Taiwan}
\affiliation[c]{Center for Theory and Computation, National Tsing Hua University, Hsinchu 300, Taiwan}

\abstract{Two classes of far detectors have been proposed or are under operation at the LHC.
The first class is a series of neutrino detectors that are sensitive to light active neutrinos via either charged-current or neutral-current interactions; exemplary ideas are FASER$\nu$, SND@LHC, and FLArE. 
Another type aims primarily at looking for displaced decays of long-lived particles (LLPs) into charged final-state particles, including ANUBIS and FASER.
In this work, we propose searches for probing lepton number violation associated with a Majorana active/sterile neutrino, for the first time with these experiments, which, if discovered, would be a clear signature of new physics beyond the Standard Model.
With Monte-Carlo simulation, we find that while the neutrino detectors, unfortunately, are estimated to have signal-event rates orders of magnitude below $\mathcal{O}(1)$, some LLP far detectors such as ANUBIS, if upgraded, would be most promising for discovering a Majorana sterile neutrino of mass $\mathcal{O}(\text{1})$ GeV in certain so-far unexcluded parameter space.
In this exploratory work, we emphasize on the importance of leveraging the LHC far detectors for purposes beyond the planned ones, such as searching for lepton number violation.
}
\begin{document}

\maketitle

\section{Introduction}\label{sec:intro}

The observation of neutrino oscillation~\cite{deSalas:2020pgw,Capozzi:2021fjo,Esteban:2020cvm} has firmly established the non-vanishing though tiny values of the light neutrino masses, providing the first evidence of physics beyond the Standard Model (BSM) and requiring extensions of the Standard Model (SM) as an explanation.
While it is possible that neutrinos are of Dirac nature implying extremely small Yukawa couplings with right-handed neutrinos, perhaps the leading candidate to explain in a natural way the tiny nonzero neutrino masses is through the dimension-5 Weinberg operator $LHLH$~\cite{Weinberg:1979sa} where lepton number is violated and the neutrinos are of Majorana nature (particles that are identical to their anti-particles).
A class of so-called ``seesaw mechanism'' models apply the latter approach, among which perhaps the type-I seesaw model is the best known~\cite{Minkowski:1977sc,Yanagida:1979as,Gell-Mann:1979vob,Mohapatra:1979ia,Schechter:1980gr}, where right-handed SM-singlet neutrinos with GUT-scale masses are introduced, predicting the tiny neutrino masses via the relation $m_\nu\sim y_\nu^2 v^2/m_N$, where $m_\nu$ is the light neutrino mass, $y_\nu$ is a Yukawa coupling, $v$ is the SM Higgs vacuum expectation value (VEV), and $m_N$ is the heavy neutrino mass.
More accommodating models include $\nu$MSM (neutrino minimal Standard Model)~\cite{Appelquist:2002me,Appelquist:2003uu,Asaka:2005an,Asaka:2005pn} which incorporates the seesaw mechanism and is able to solve additional issues of the SM such as dark matter~\cite{Dodelson:1993je,Shi:1998km} and baryon asymmetry~\cite{Asaka:2005an,Asaka:2005pn}.

An important question hence arises: are neutrinos Dirac or Majorana particles?
To answer the question, one of the main methods is to search for lepton-number-violating (LNV) signal processes.
For instance, observation of neutrinoless double beta decay of atomic nuclei~\cite{Furry:1939qr} has been pursued since decades ago, and if made, would be direct evidence for LNV by two units, indicating the neutrinos' Majorana nature~\cite{Schechter:1981bd,Duerr:2011zd}.
Recent reviews can be found in Refs.~\cite{Rodejohann:2011mu,DellOro:2016tmg,Dolinski:2019nrj}.
LNV processes can also be searched for in the energy frontier, e.g.~at the LHC~\cite{Maiezza:2015lza,ATLAS:2018dcj,CMS:2018jxx,LHCb:2014osd}.
If observed, the Majorana nature of neutrinos can be determined~\cite{Babu:2022ycv}.
One of the first LHC search strategies proposed for probing LNV is the famous Keung-Senjanovic process~\cite{Keung:1983uu}, $pp\to l^\pm l^\pm jj$, studied in the context of Left-Right Symmetric Model with a right-handed $W_R$ boson and right-handed neutrinos $N$~\cite{Pati:1974yy,Mohapatra:1974gc,Mohapatra:1980yp}.
In the final states, there are same-sign leptons and fully reconstructed hadronic activities, allowing to establish $|\Delta L|=2$ when compared to the initial-state lepton number $L=0$.
Searches for the LNV signals have been performed, e.g.~at ATLAS~\cite{ATLAS:2018dcj} and CMS~\cite{CMS:2018jxx} for $W$-boson decays, as well as at LHCb~\cite{LHCb:2014osd} where $B$-meson decays were considered.
Finally, long-baseline neutrino experiments can also be used for searching for LNV phenomena, e.g.~in neutrino oscillations~\cite{Bolton:2019wta}.
See Refs.~\cite{Deppisch:2015qwa,Cai:2017mow} for recent reviews on neutrinos, lepton number violation, and collider physics.

However, to confirm the LNV, previous collider experiments require that the new particles (e.g.~the heavy neutrino) must decay inside the near detector.
If the particle is stable (e.g.~the active neutrino), or the lifetime of the new particle is very long (e.g.~the heavy neutrino with very small mass and tiny active-heavy neutrino mixings), once produced, it escapes the near detector behaving as missing energy.
As a result, their lepton number is not measured, and thus the LNV cannot be confirmed for these particles by the previous collider experiments.
To ensure observation of LNV, the lepton numbers for such stable or ``nearly" stable particles must be measured.
Although this cannot be realized with the near detector, it can be achieved with the recently installed or proposed neutrino far detectors at the LHC, where neutrinos are detected and their lepton numbers are measured through the neutrino-nucleus charged-current deep inelastic scattering.
Such ongoing and future experimental efforts on neutrino physics include FASER$\nu$~\cite{FASER:2019dxq,FASER:2020gpr} and SND@LHC~\cite{
SHiP:2020sos,SNDLHC:2022ihg} currently operated during Run 3, and FASER$\nu$2~\cite{MammenAbraham:2020hex,Anchordoqui:2021ghd,Feng:2022inv}, AdvSND~\cite{Feng:2022inv,MammenAbraham:2022xoc}, and FLArE-10/100~\cite{Feng:2022inv,Batell:2021blf} to be running during the high-luminosity LHC (HL-LHC) era at the proposed Forward Physics Facility (FPF)~\cite{Feng:2022inv}.
These experiments are composed of dense materials such as tungsten or liquid argon as target, installed all in the very forward direction of the ATLAS interaction point (IP) about 400-600 meters away, and are able to detect the neutrinos and measure their lepton numbers through the neutrino-nucleus deep inelastic scattering.
Despite the relatively small interaction probabilities, it has been estimated that the huge numbers of neutrinos produced mainly from meson decays would still allow for, e.g.~observation of up to tens of thousands of neutrinos at FASER$\nu$~\cite{FASER:2019dxq}.
In fact, the first observation of these neutrinos has already been made at FASER$\nu$~\cite{FASER:2021mtu,FASER:2023zcr}.

Here, we ask the question: how to utilize these neutrino far detectors to confirm LNV signals and discover Majorana particles, especially for the stable or ``nearly" stable particles?
To answer the question, in this study, we focus on neutrinos from leptonic decays of the $W$-bosons at the IP, and require event correlation between the near detector and the neutrino far detector.
This channel gives transversely moving charged leptons with high $p_T$, rendering the signature particularly clean.
By measuring the charge sign of the prompt lepton and the charged lepton produced from neutrino-nucleus scattering, one can confirm observation of same-sign leptons; since no missing energy exists in the complete event, LNV can thus be established.
Note that, in this study, we assume that these neutrino far detectors are able to determine the electric sign of charged leptons, and to measure the 4-momentum of the outgoing charged lepton and the nuclear recoil.
Besides, for simplicity, we will restrict ourselves to active neutrinos of the electron or muon flavor.

For this search, the main limiting factor is the strong helicity flip suppression that would be relevant for the Majorana light neutrinos.
For purely $W$-boson charged-current (CC) interactions, such a suppression factor is about $m^2_\nu/(4 E^2_\nu)\sim 2.5\times 10^{-25}$ for neutrino mass $m_\nu=0.1$ eV and a typical energy $E_\nu$ of 100 GeV at the LHC.
A simple order-of-magnitude estimate taking into account only the integrated luminosity, active neutrino production cross section, as well as the helicity flip, can be performed.
The production cross section of one species of active neutrinos from $W^\pm$-bosons decays at the LHC is about $20$ nb~\cite{ATLAS:2016fij}.
With an integrated luminosity $\mathcal{L}$ of $3$ ab$^{-1}$ expected at the end of the high-luminosity LHC, we find the expected signal event rates are already only at the order of $10^{-14}$, assuming $100\%$ acceptance and efficiencies, as well as the scattering interaction probabilities.
Nevertheless, in this work, we perform a detailed analysis of this scenario where we take into account the detector acceptances and the scattering interaction probabilities, in order to serve several purposes.
Firstly, we develop the search strategy and establish the analysis techniques which can be used for other scenarios similar to those studied in this work.
Secondly, the full calculation allows to determine quantitatively the improvement that would be required in order to be able to observe signal events.
Finally, for potential future experiments designed for similar research aims, our computation provides a baseline benchmark hinting at possible experimental upgrades that should be implemented at priority.

We do not consider active neutrinos from  meson decays, because these mesons and hence the charged leptons that they decay to are mostly soft and traveling in the forward direction.
Since the prompt charged leptons travel with a large pseudorapidity and a small transverse momentum, they would be almost impossible to be detected by the near detector.
Thus, their lepton numbers cannot be measured.
In other words, it would be difficult to trigger the signal event at the near detector and at the same time to have an energetic neutrino from meson decays travel in the very forward direction hitting the neutrino far detectors.
Moreover, even in the forward regime at ATLAS, hundreds or even up to thousands of tracks exist, originating from e.g.~pile-up effects, multi-parton interactions, as well as proton beam remnants.
This makes it also extremely hard to ensure the prompt and scattered charged leptons originate from the same collision event (event correlation).
As a result, it would be highly unlikely, if not impossible, to observe same-sign leptons and hence LNV.

We emphasize again that such a search for same-sign leptons is impossible to perform with the current experiments or proposals.
For instance, the ATLAS or CMS near detectors solely cannot detect the active neutrinos from $W$-boson decays leading to transverse missing momentum; therefore, the lepton number in the final state cannot be determined.
Moreover, the existing and future neutrino far detectors mainly study neutrinos from meson decays; they do not require event correlation between the IP and the neutrino detector, nor is it possible to decide the charge sign of the initial-state leptons from meson decays.
For these reasons, these experiments cannot confirm observation of LNV.
We will discuss the detail of our search in Sec.~\ref{sec:exp}.

Almost at the same location of these neutrino detectors, a different series of far-detector programs have also been proposed to be operated during Run 3 or HL-LHC.
Focusing on tracks, these experiments will mainly search for displaced vertices stemming from decays of long-lived particles (LLPs) that are predicted in many BSM models, such as sterile neutrinos, dark photons, and dark Higgs bosons (see Refs.~\cite{Alimena:2019zri,Lee:2018pag,Curtin:2018mvb,Knapen:2022afb,Blondel:2022qqo} for recent reviews on LLPs).
For example, FASER~\cite{Feng:2017uoz,FASER:2019aik,FASER:2022hcn} is a small cylindrical detector installed right behind FASER$\nu$ and is able to probe a wide range of BSM scenarios with LLPs~\cite{FASER:2018eoc}.
Alternatively, in one of the service shafts above the ATLAS IP, another idea called ANUBIS~\cite{Bauer:2019vqk} has also been suggested to be placed, which has been predicted to have strong sensitivities to various types of LLPs~\cite{Hirsch:2020klk,Dreiner:2020qbi,DeVries:2020jbs,Cottin:2021lzz,Mitsou:2020okk,Mitsou:2021tti,Bertuzzo:2022ozu,Barducci:2022gdv}.
Near the vicinity of the CMS IP, similar concepts such as MATHUSLA~\cite{Curtin:2018mvb,Chou:2016lxi,MATHUSLA:2020uve} and FACET~\cite{Cerci:2021nlb} exist, too.
We will consider long-lived Majorana sterile neutrinos produced from $W$-boson decays at the IP, and then decaying semi-leptonically in the LLP far detectors.
For simplicity, we focus on the case that there is only one kinematically relevant sterile neutrino and it mixes with the electron neutrino only.
Studies on searches for such long-lived sterile neutrinos at the proposed far detectors have been extensively published (see e.g.~Refs.~\cite{Helo:2018qej,Curtin:2018mvb,Kling:2018wct,Hirsch:2020klk}).
These searches usually assume zero background and estimate the inclusive number of displaced vertices of the HNLs inside the LHC far detectors such as FASER and MATHUSLA, dominantly produced from rare decays of kaons, $D$-mesons, and $B$-mesons.
Since they do not decide the lepton number in these meson decays, nor do they impose event correlation requirement, these searches are usually unable to look for same-sign leptons and hence to determine LNV in the whole event process.
As explained in the previous paragraph, the final-state particles of these meson-decay processes are soft and mostly travel in the very forward direction, making it unrealistic to pin down the prompt charged lepton at ATLAS or CMS.
Therefore, we have chosen to focus on $W$-boson decays, too, for looking for LNV signatures with the present search strategy.
We further assume the sterile neutrino mass $m_N$ and the mixing angle $V_{eN}$ as two independent parameters for the purpose of phenomenological studies.
Similar to the case of neutrino detectors, the search strategy requires that the LLP far detector should have event correlation with the prompt activities, and be able to measure the sign of charged leptons and the 4-momenta of final state particles\footnote{In Ref.~\cite{Dib:2015oka} it was pointed out that for sterile neutrinos produced from $W$-bosons and decaying inside the near detector such as ATLAS, one can distinguish between the Dirac and Majorana sterile neutrinos by studying the final-state lepton's spectrum. However, this requires at least a reasonably large event sample. See also Ref.~\cite{Dib:2016wge}.}.
Detailed discussion is provided in Sec.~\ref{sec:exp}.

Following Sec.~\ref{sec:exp}, we present numerical results in Sec.~\ref{sec:results}, for neutrino-nucleus scattering first and then displaced decays of long-lived sterile neutrinos.
Finally, we summarize the work in Sec.~\ref{sec:conclusions}.
Additionally, a detailed discussion on the use of leptoquark models for emulating dimension-7 LNV effective-field-theory operators is provided in Appendix~\ref{app:LQ}.

\section{Experiments and the proposed searches}\label{sec:exp}

In this section, we discuss the different types of experiments we consider.
We will first cover the LHC neutrino detectors, explaining how to estimate the total signal-event numbers in detail.
We will then turn to the LHC LLP far detectors, elaborating on the computation procedure of the signal-event rates for our search strategy.
We note that since these searches are expected to be background-free, observation of a few events should already be sufficient for establishing the discovery of LNV and a Majorana neutrino.
\subsection{Neutrino detectors}\label{subsec:v_detectors}

\begin{table}[t]
\centering
\begin{tabular}{c|cccccc}
\hline
\hline
neutrino detectors                                                                 & material                                               & $A$ [cm$^2$]                          & $m_{\text{det}}$ [ton]       & $\eta_{\text{min}}$     & $\eta_{\text{max}}$          & $\mathcal{L}~[\text{fb}^{-1}]$ \\ \hline
FASER$\nu$~\cite{FASER:2020gpr,Kling:2021gos,Kling:2020iar}      & tungsten                                               & $25\times 25$                          & 1.2                       & 8.5                     & $\infty$                     & 150                         \\ 
SND@LHC~\cite{SHiP:2020sos,SNDLHC:2022ihg}                       & tungsten                                               & $40\times 40$                          & 0.85                        & 7.2                     & 8.4                          & 150                         \\
\hline
FASER$\nu$2~\cite{Feng:2022inv}                                  & tungsten                                               & $40\times 40$                          & 20                      & 8.5                     & $\infty$                     & 3000                        \\
AdvSND(far)~\cite{Feng:2022inv,MammenAbraham:2022xoc}            & tungsten                                               & $100\times 55$     & 5                       & 7.2                     & 8.4                          & 3000                        \\ 
FLArE-10~\cite{Batell:2021blf,Feng:2022inv}                      & liquid argon                  & $100 \times 100$                       & 10                      & 7.5 & $\infty$                     & 3000                        \\ 
FLArE-100~\cite{Batell:2021blf,Feng:2022inv}                      & liquid argon                     & $160 \times 160$                       & 100                    & 7 & $\infty$                     & 3000              \\
\hline
\hline      
\end{tabular}
\caption{Summary of neutrino detectors at the LHC. We list the material type, area $A$, detector mass $m_{\text{det.}}$, minimal and maximal pseudorapidity coverage $\eta_{\text{min}}$ and $\eta_{\text{max}}$, as well as the corresponding integrated luminosity, $\mathcal{L}$, for each detector.}
\label{tab:v_exp}
\end{table}

We give a brief discussion on a series of neutrino detectors proposed in the forward direction of the LHC ATLAS IP.
During LHC Run 3, FASER$\nu$ and SND@LHC are collecting data, studying active neutrinos.
Both consist of a tungsten target of mass about one ton, and cover slightly different pseudorapidity ranges.
Although the LHC Run 3 schedule has been adjusted as a result of the COVID-19 pandemic and it is now slated to accumulate 300 fb$^{-1}$ integrated luminosity, in this work we stick to 150 fb$^{-1}$ as a benchmark value.
At the proposed program FPF, several more neutrino detectors are now in plan.
A successor of FASER$\nu$, FASER$\nu$2 would apply the same target material, but with a larger volume and hence larger weight, up to 20 tons.
Similarly, SND@LHC has a follow-up upgrade experiment called AdvSND.
It consists of both near and far detectors; here, we will study the far detector, which would have 5 tons of tungsten material.
At the end, FLArE may employ a target of either 10 tons or 100 tons of liquid argon material.
These proposals at FPF would be working during the HL-LHC era, corresponding to about 3 ab$^{-1}$ integrated luminosity by the end of the HL-LHC schedule.
All these detectors are in the forward direction, as a huge number of neutrinos are produced from meson decays and traveling in this direction.
A summary of these experiments' information can be found in Table~\ref{tab:v_exp}.

When neutrinos hit these targets, they can scatter with the nucleons in the nuclei, resulting in either neutral-current (NC, via a $t$-channel $Z$-boson in the SM) or charged-current (CC, via a $t$-channel $W$-boson in the SM) interactions, which can be detected by nuclear recoil~\cite{Ismail:2020yqc,Cheung:2021tmx} or observing the produced outgoing charged lepton (\cite{Ansarifard:2021dju,Kling:2020iar}), respectively.
In addition, such signatures can also arise with higher-dimensional operators involving two quarks and two neutrinos (or one neutrino and one charged lepton).
We will consider both possibilities and explore their differences, as explained in Sec.~\ref{sec:v}.

In this work, we focus on $W$-boson decays at the ATLAS IP for the light neutrino production, in association with a prompt charged lepton.
The prompt lepton serves as a trigger for a mono-lepton event, given that it is sufficiently ``hard''.
Fixing the active neutrino mass to be 0.1 eV in accordance with neutrino oscillation~\cite{Canetti:2010aw}\footnote{For more recent global analyses, see Refs.~\cite{Capozzi:2021fjo,Esteban:2020cvm,deSalas:2020pgw}.} and cosmological constraints~\cite{Planck:2018vyg}, we perform Monte-Carlo (MC) simulation with the tool \texttt{Pythia8.308}~\cite{Sjostrand:2014zea,Bierlich:2022pfr}, generating 10 million $W$-bosons from $pp$ collisions which subsequently decay to an electron or a muon, plus a light neutrino.
If the produced light neutrino travels in the direction of the neutrino detectors, it can scatter with the target, producing a charged lepton.
In order to probe LNV, the four-momenta of both the prompt lepton at the IP and the outgoing lepton at the neutrino detectors need to be measured.
Moreover, the nuclear recoil should be measured.
These are to ensure the non-existence of missing energy (in the transverse direction).
Event correlation between the prompt and scattering leptons should be realized; this is challenging, though not impossible.
For instance, even though FASER$\nu$ as an emulsion detector is not equipped with timing capabilities, an interface tracking layer has been added to connect FASER$\nu$ to FASER~\cite{FASER:2022hcn}, enabling to record the time of the scattered muons.
Similar concepts of timing layers have also been proposed for FLArE~\cite{Batell:2021blf} and SND@LHC~\cite{SHiP:2020sos}.
At the end, we require that the prompt and the scattered leptons should have the same electric charge sign; this can be measured only if magnets are installed at the neutrino detectors.
For example, FASER$\nu$ is connected to the FASER magnetic spectrometer, and is hence able to discern between positively and negatively charged particles.
Further, muons and electrons can be differentiated through their track lengths~\cite{FASER:2019dxq}.
Similarly, the most downstream element of AdvSND is a magnet enabling muon charge and momentum measurement\cite{Feng:2022inv}.
For FLArE, it is still under discussion regarding the installation of magnetic field and a potential magnet or a time projection chamber for momentum measurement~\cite{Feng:2022inv}.

\begin{figure}[t]
    \centering
    \includegraphics[width=0.495\textwidth]{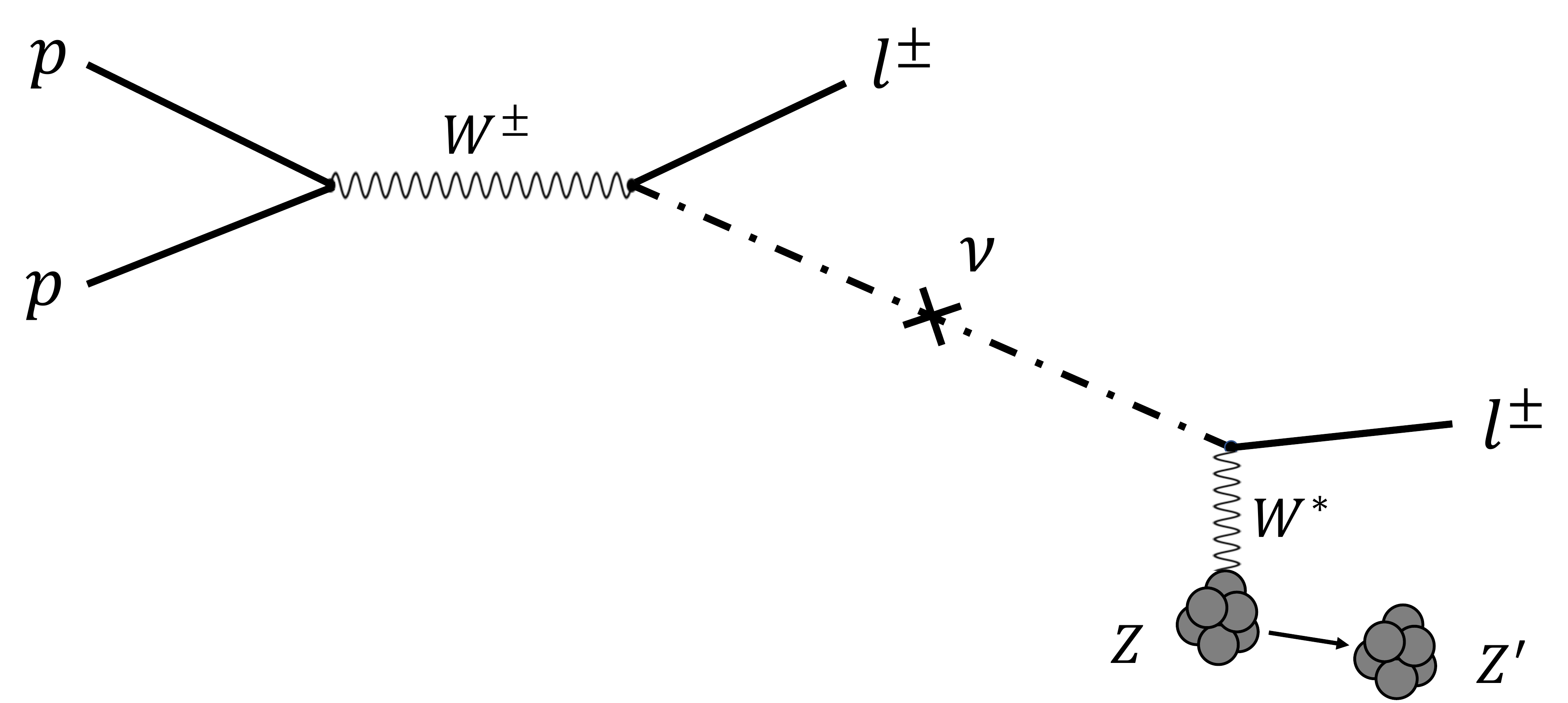}
    \includegraphics[width=0.495\textwidth]{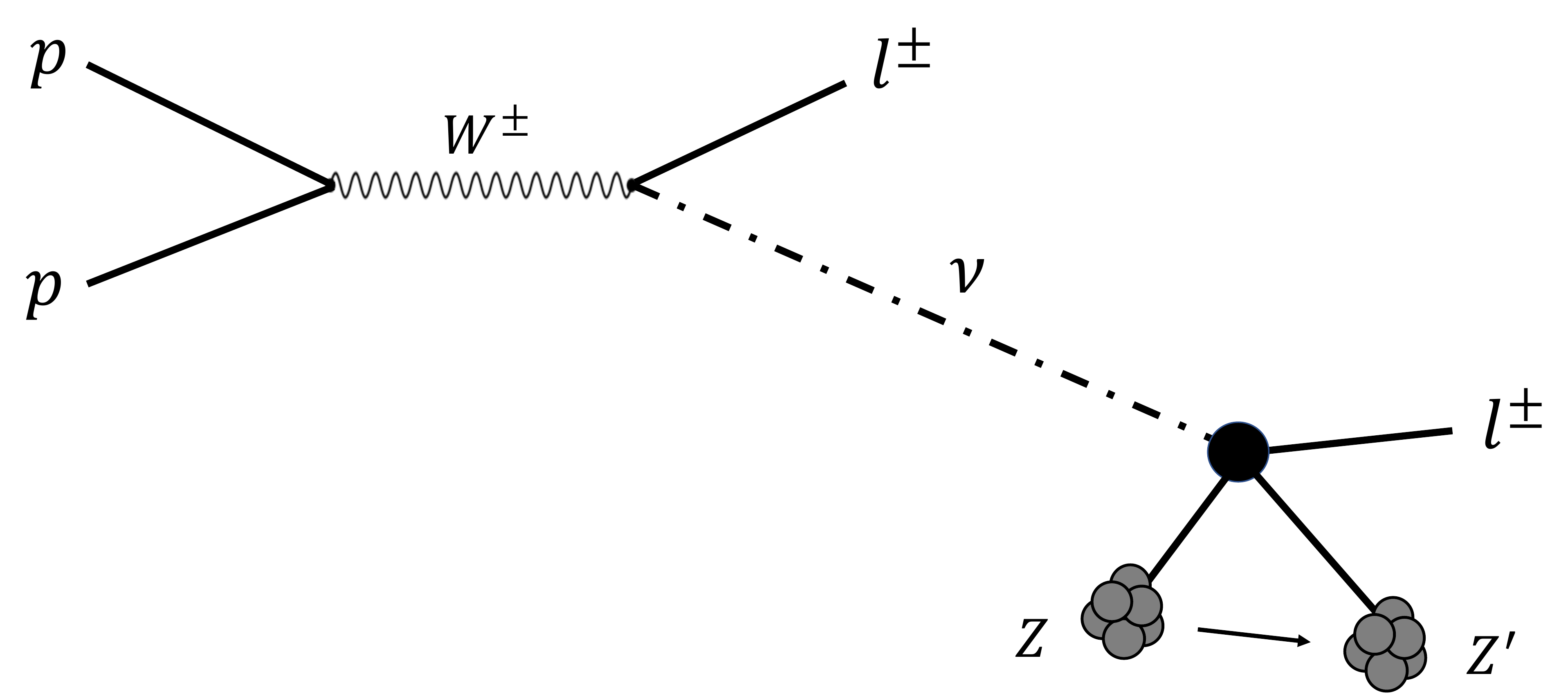}\\
    \includegraphics[width=0.5\textwidth]{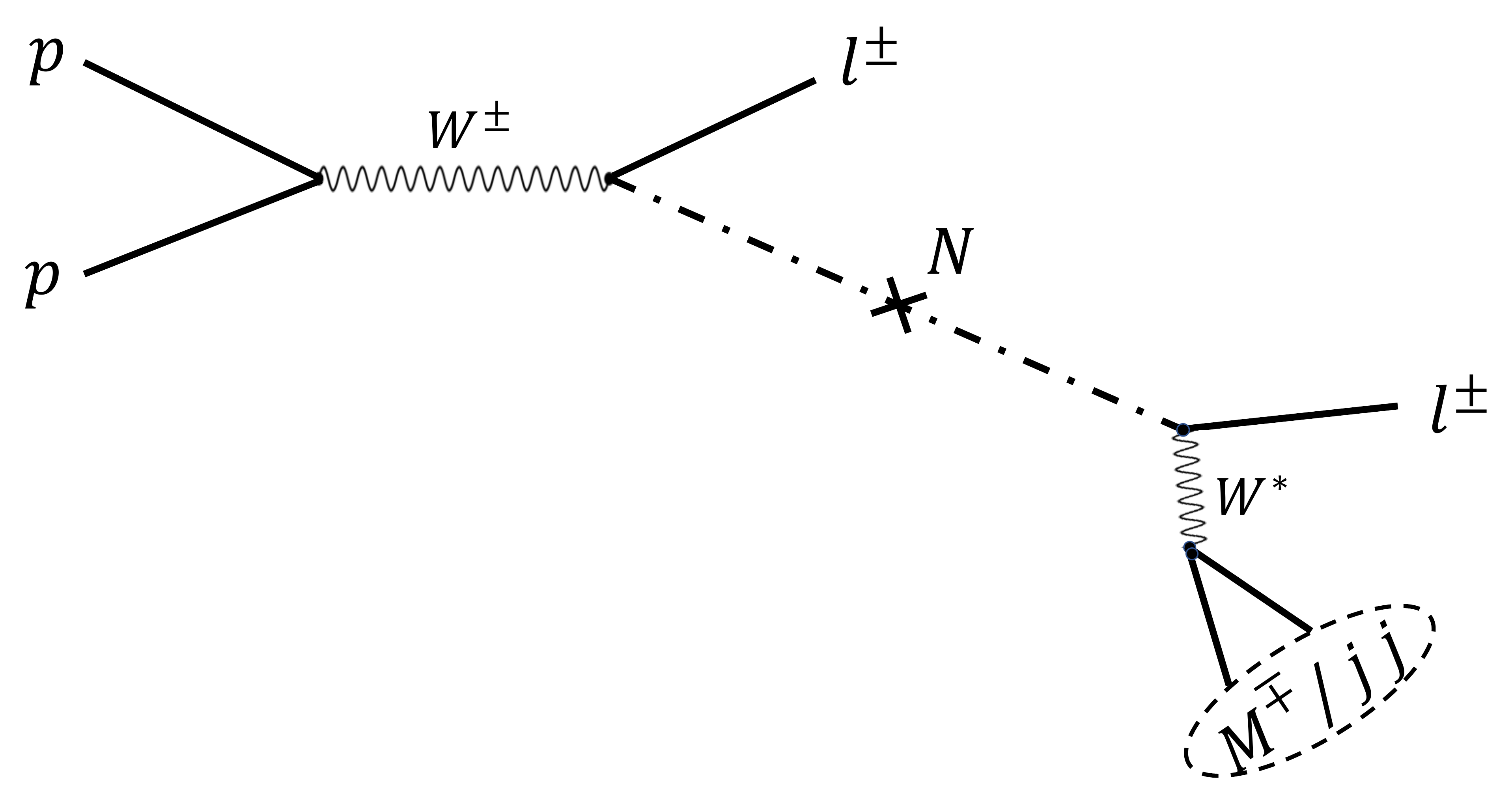}
    \caption{Sketches of the signal processes for the neutrino-nucleus scattering case (upper panel) and the sterile neutrino displaced decay case (lower panel). $Z$ and $Z'$ label the nucleus, and the solid black bulb in the upper right plot denotes the effective vertex of the dim-7 LNV operator. The cross on the neutrino line in the upper left plot corresponds to the helicity flip required for achieving LNV in this case.}
    \label{fig:sProc}
\end{figure}

The experimental total cross sections of active neutrino production from $W$-boson decays ($pp\to W^\pm \to l^\pm \nu$), $\sigma_{\nu}$, have been measured at ATLAS~\cite{ATLAS:2016fij}.
The efficiency of the prompt leptons passing the trigger requirement $\epsilon_{\text{trigger}}$ can be retrieved from \texttt{Pythia8} simulation, where the trigger requirement is defined as follows:
$p^e_T>27.3$ GeV and $|\eta^e|<2.5$ for electrons from the $W$-boson decay, and $p_T^\mu>27$ GeV and $|\eta^\mu|<2.5$ for muons~\cite{ATLAS:2020xea,ATLAS:2022atq}.
We further define $\epsilon_{\text{window}}$ as the efficiency of the light neutrino from the $W$-boson decay traveling inside the neutrino detector window, after the event passes the trigger requirement; $\epsilon_{\text{window}}$ is also obtained from simulation.
The interaction probability of a neutrino with the detector can be computed with the following formula~\cite{FASER:2019dxq}:
\begin{eqnarray}
P_{\text{scatt.}} = \frac{\sigma_{\nu Z}}{A}\frac{m_{\text{det}}}{m_Z},
\label{eq:Pscatt}
\end{eqnarray}
where $\sigma_{\nu Z}$ is the CC scattering cross section and is a function of the neutrino energy, $A$ is the detector area perpendicular to the beam direction, $m_{\text{det}}$ is the detector total mass, and $m_Z$ is the nucleus mass.
We confine ourselves to neutrino energies between 10 GeV and 10 TeV for neutrino-nucleus scattering at these neutrino detectors.
The lower energy threshold is chosen because for $E_\nu\lesssim 10$ GeV, the scattering is dominated by quasi-elastic scattering and resonant production processes, and the upper energy threshold 10 TeV corresponds to $Q^2\sim m_W^2$ and roughly the maximal possible neutrino energy at the LHC.
For the SM CC, $\sigma_{\nu Z}$ has been computed in Ref.~\cite{FASER:2019dxq} and will be extracted therefrom, and for the higher-dimensional effective operators we will compute the CC scattering cross sections ourselves.
At the end, the signal event rate $N_S^\nu$ can be estimated with
\begin{eqnarray}
    N_S^\nu = \sigma_{\nu}\cdot \mathcal{L} \cdot \epsilon_{\text{trigger}}\cdot \epsilon_{\text{window}} \cdot \langle  \epsilon_{\text{h.~f.}} \cdot P_{\text{scatt.}} \rangle,
    \label{eq:NSv}
\end{eqnarray}
where $\mathcal{L}$ is the integrated luminosity.   
Further, $\langle \epsilon_{\text{h.~f.}} \cdot P_{\text{scatt.}}\rangle$ labels the average value of the product of the helicity flip suppression factor $\epsilon_{\text{h.~f.}}$ and the scattering interaction probability of the neutrinos that have passed both the prompt-lepton ``trigger'' and the neutrino detector ``window'' requirements, obtained through the \texttt{Pythia8} MC simulation.
For $\epsilon_{\text{h.~f.}}=m^2_\nu/(4 E^2_\nu)$, a simple estimate has been made in Sec.~\ref{sec:intro} and therefore helicity flip suppression has been well explained for the purely $W$-boson CC-interaction scenario.
However, if either the production or the scattering interaction, but not both, is itself LNV, such a helicity flip should not be included, i.e.~$\epsilon_{\text{h.~f.}}=1$, for achieving LNV in the whole process.
Accordingly, in this work, we will consider an additional case, where the production is via the SM CC while the scattering is mediated by a dim-7 LNV operator.
We will discuss these in more detail in Sec.~\ref{sec:v}.
A simple sketch of the two neutrino-scattering signal processes is shown in the upper plots of Fig.~\ref{fig:sProc}.
The difference between the two plots lies mainly in the interaction responsible for the neutrino-nucleus scattering, being due to the SM charged current or a higher-dimensional effective operator.
Finally, we note that we consider $100\%$ detection efficiency for electrons and muons at these neutrino detectors.

\subsection{LLP far detectors}\label{subsec:LLP_detectors}

We discuss the other class of experiments which are designed mainly for searching for displaced vertices from LLPs (``LLP far detectors'').
These include various far-detector experiments proposed in the vicinity of ATLAS, CMS, or LHCb IP, with a distance of about 10 to 500 meters.
For the ATLAS IP, FASER is a small cylindrical detector that has been installed in the very forward direction with a distance of 480 m from the ATLAS IP and is already collecting data during Run 3.
FASER2 is a follow-up program of FASER, to be installed at FPF with a distance of 620 m from the ATLAS IP.
Further, in a service shaft above the ATLAS IP, a larger detector called ANUBIS would be constructed; it also has a cylindrical shape but faces vertically.
Then in the forward direction of the CMS IP, a CMS sub-system called FACET has been brought up to be placed surrounding the the beam pipe.
Moreover, in the transverse side of the CMS IP, a huge detector, MATHUSLA, would be installed, about two hundred meters away from the IP.
Finally, for the LHCb IP, some far-detector proposals currently exist: CODEX-b~\cite{Gligorov:2017nwh,Aielli:2019ivi}, and MoEDAL-MAPP1 and MAPP2~\cite{
Pinfold:2019nqj,Pinfold:2019zwp}, among which MAPP1 is under operation at the moment during Run 3.
For a summary of these detectors including their geometries and corresponding integrated luminosities, see e.g.~Refs.~\cite{DeVries:2020jbs,Cerci:2021nlb}.

The LHCb detector has an acceptance for prompt leptons that have a pseudorapidity between 2 and 5~\cite{LHCb:2008vvz}.
The accompanying sterile neutrino from $W$-boson decays also tends to travel in this direction.
However, CODEX-b is in the $[0.2,0.6]$ pseudorapidity range and the MoEDAL-MAPP detectors are even in the negative pseudorapidity hemisphere.
Therefore, neither CODEX-b nor MAPP1(2) would be receiving the sterile neutrinos produced from $W$-boson decay events for which the trigger requirement on the prompt lepton is already satisfied.
Further, the two experiments using events from the CMS IP have intrinsic deficiencies.
To determine the displaced lepton charge sign, it is required to install magnets around the far detectors\footnote{Since it would still be difficult to distinguish a muon and a charged hadron with magnets and trackers, we would restrict ourselves to the case that the sterile neutrino mixes with the electron active neutrino only.}.
Given the gigantic volume of MATHUSLA that is more than 100 thousand m$^3$, it is virtually impossible to do so, given the huge ensuing cost.
As for FACET, magnets cannot be placed there since they would affect the LHC proton beams that are supposed to travel inside the beam pipe under the influence of the existing superconducting magnets.

At the end, for FASER(2) and ANUBIS, these issues do not exist or are not so severe.
Sec.~6.5 of Ref.~\cite{Anchordoqui:2021ghd} has discussed the feasibility of correlating events from ATLAS and from an experiment at the FPF such as FASER2, which would primarily depend on the possibility of triggering ATLAS by a FASER2 event, and we summarize the main conclusions here.
First, it requires a timing resolution better than 25 ns in order to associate a FASER2 signal event to a certain bunch crossing at the ATLAS IP.
Further, for a distance of approximately $500-600$ m of the FASER2 detector from the ATLAS IP, taking into account the trigger latency of about \SI{10}{\micro\second} for the ATLAS Level-0 system and the time required for a light GeV-scale HNL to reach FASER2 and for the trigger signal to arrive at the Central Trigger Processor, the triggering should take place within $5-6$ \SI{}{\micro\second} so that it can be used by the ATLAS Level-1 trigger system.
A second issue is the many pileup events at the HL-LHC.
A timing resolution of 100 ps would be required to associate a forward signal event to a part of the luminous region where the particles in the forward signal originated, so as to reduce the pileup background.
In the present days a timing resolution of $\mathcal{O}(1\text{ ns})$ is within reach, while future technical development would be required to achieve timing resolutions of about $100$ ps. 
While FASER is not equipped with such timing resolution abilities, similar arguments can still be applied for the ANUBIS experiment, where, though, because of the closer distance from the IP, the triggering should be decided even faster (see also Appendix A of Ref.~\cite{Cottin:2021lzz} for a discussion on the effect of a timing cut (removing time-delayed signal events in order to reduce background events) at ANUBIS on LLP sensitivities, which would be mainly relevant for LLPs heavier than $\sim 100$ GeV).
We note that for our theoretical scenario where the HNL mass is around the GeV scale, the HNLs travel almost at the speed of light and therefore, an issue that could arise from a time-delayed signal is absent here.
Once FASER2 and ANUBIS are equipped with the above discussed timing capabilities, it should be technically possible to realize the event correlation required by our proposed search.
For numerical results to be presented in the next section, we will nevertheless show sensitivity curves for not only FASER2 and ANUBIS, but also FACET and MATHUSLA.
The whole process of the signal event is shown in the lower plot of Fig.~\ref{fig:sProc}.

To compute the LNV sensitivities of these far detectors, we focus on $W$-boson decays into an electron plus a sterile neutrino that mixes only with the electron neutrino since it would be difficult to distinguish a muon from a charged hadron with the proposed hardware upgrades.
As in the neutrino-scattering study, we use \texttt{Pythia8} to obtain the trigger efficiency $\epsilon_{\text{trigger}}$ for the prompt charged leptons.
For events passing the trigger requirement, we calculate the average decay probabilities of the sterile neutrinos inside the far detectors, $\langle P_{\text{decay}} \rangle$, with exponential decay laws; the exact formulas for $P_{\text{decay}}$ depend on the sterile neutrinos' lifetime, speed, and traveling direction, as well as the detectors' geometries, and can be found in Ref.~\cite{DeVries:2020jbs} and the references therein.
The required kinematic information can be extracted from the \texttt{Pythia8} simulation.
Thus, we can estimate the total signal-event number $N_S^N$ with
\begin{eqnarray}
    N_S^N = \frac{\sigma_\nu}{\text{BR}(W\to e\, \nu)}\cdot \text{BR}(W^\pm \to e^\pm \, N) \cdot \mathcal{L}\cdot \epsilon_{\text{trigger}} \cdot \langle P_{\text{decay}} \rangle  \cdot \text{BR}(N\to e^\pm  jj/M^\mp ),\,\,\,
    \label{eq:NSN}
\end{eqnarray}
where the $W$-boson decay branching fraction into an electron and a sterile neutrino is computed with
\begin{eqnarray}
    \text{BR}(W^\pm\to e^\pm\, N) = \frac{1}{\Gamma_W}\,\frac{G_F}{\sqrt{2}}\frac{m_W^3}{12 \pi} |V_{eN}|^2 \Big(  2 +      \frac{m_N^2}{m_W^2}  \Big) \cdot \Big(  1 - \frac{m_N^2}{m_W^2}  \Big)^2,
\end{eqnarray}
with $\Gamma_W=2.085$ GeV denoting the $W$-boson total decay width~\cite{ParticleDataGroup:2020ssz}, $G_F$ the Fermi constant, $m_W$ the $W$-boson mass, $|V_{eN}|^2$ the active-sterile neutrino mixing, and $m_N$ the sterile neutrino mass.
We stress here, that one advantage of this displaced-vertex search compared to the LNV neutrino-scattering searches associated with a Majorana active neutrino discussed above is the absence of the (strong) helicity flip suppression effect despite the existence of helicity inversion, as a result of the on-shellness of the HNL as well as its narrow decay width~\cite{Ruiz:2020cjx}.
Finally, we note that for $m_N\gtrsim 1$ GeV, we focus on $N$ decays to the $ejj$ final states including two jets.
For $m_N \lesssim 1$ GeV, $N$ decays to a lepton and a charged meson ($M^\mp$ in Eq.~\eqref{eq:NSN})~\cite{Bondarenko:2018ptm,DeVries:2020jbs}.
Restricting ourselves to these channels allows for full reconstruction of the displaced-decay processes.

\section{Numerical results}\label{sec:results}

We present numerical results in this section.
For both types of signatures studied in this work, we expect vanishing irreducible background.
Therefore, the $2\sigma$ ($95\%$ confidence level) sensitivity bounds correspond to 3 signal events, determined by setting zero background and observed events in the Feldman-Cousins approach~\cite{Feldman:1997qc,Hill:2002nv} following Poisson distribution.

\subsection{Active neutrinos}\label{sec:v}

We begin the subsection with discussing active neutrino-nucleus scattering with the SM weak-current interactions only.
The active neutrinos are produced from $W$-boson decays and subsequently scatter with the target nucleus at the neutrino detectors into a charged lepton $e$ or $\mu$ with the same charge sign as that of the prompt lepton: $pp\to W^\pm \to l^\pm \nu, \nu Z \to l^\pm Z'$, where $Z$ and $Z'$ denote the nucleus before and after the scattering, respectively.
The $W^+$ and $W^-$ production cross sections are 108.93 nb and 80.94 nb, respectively~\cite{ATLAS:2016fij,ParticleDataGroup:2020ssz}.
As mentioned above, we simulate $10^7$ events for each neutrino detector, where about $42\%(58\%)$ is for $W^-(W^+)$ production.

\begin{figure}[t]
\centering
 \includegraphics[width=0.495\textwidth]{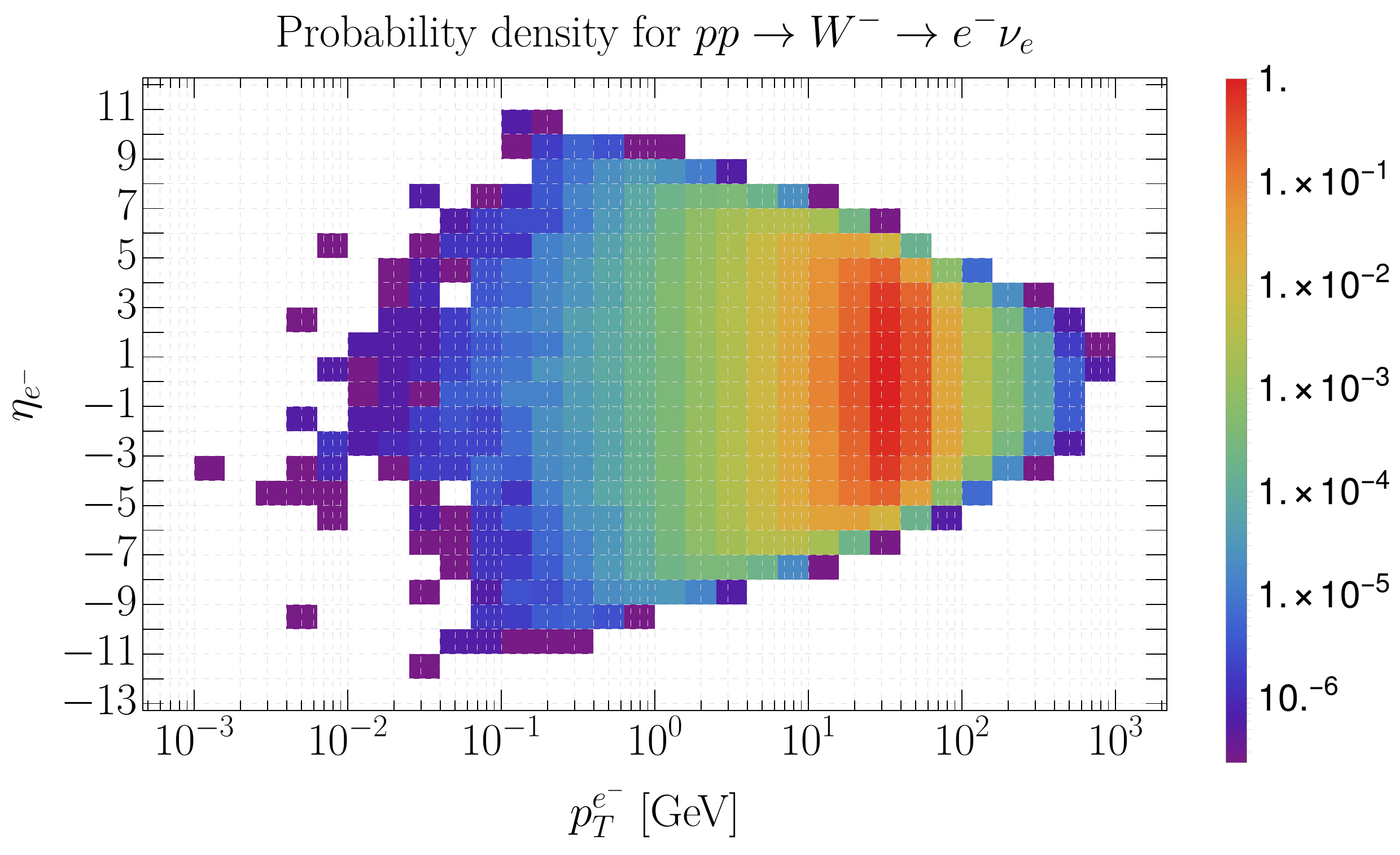}
 \includegraphics[width=0.495\textwidth]{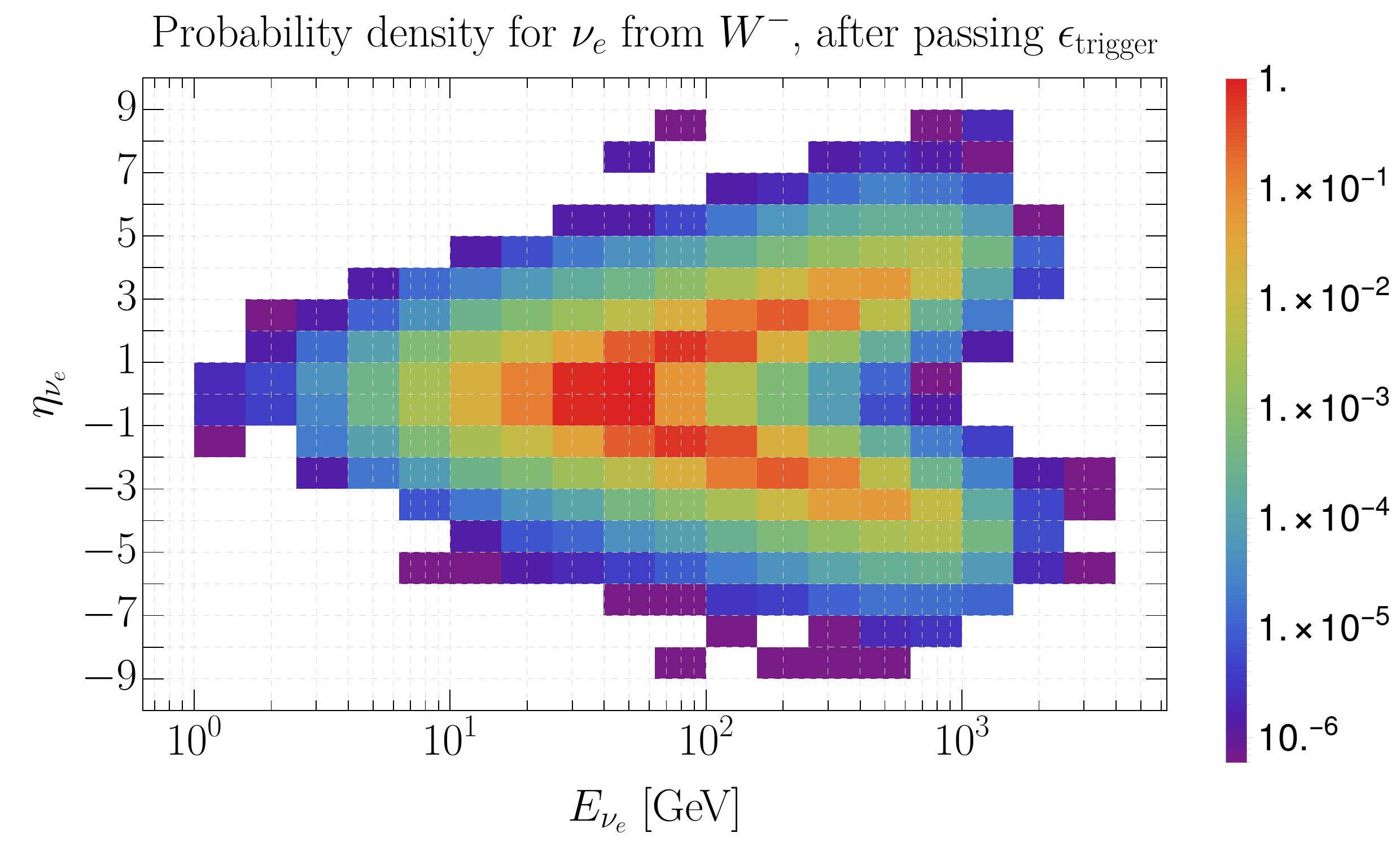}
  \caption{Normalized kinematical distributions of the prompt electron before applying the trigger requirement, and of the prompt neutrinos after applying the trigger requirement on the prompt electrons. The considered process is $pp\to W^-\to e^- \nu_e$. For the other processes where $e^+$, $\mu^-$, or $\mu^+$ is produced, the distributions are similar. $10^7$ signal events were generated with \texttt{Pythia8}.}
  \label{fig:kinematical_distributions}
\end{figure}
In the left panel of Fig.~\ref{fig:kinematical_distributions}, we present the normalized kinematical distributions of $e^-$ in the $\eta_{e^-}$ vs.~$p_T^{e^-}$ plane, produced in $pp\to W^-\to e^- \nu_e$, before applying the trigger requirement.
We find the majority of events have a transverse momentum larger than 27.3 GeV, centered around $\eta_{e^-}=0$.
For positrons from $W^+$ decays, as well as muons and anti-muons produced in the similar ways, we do not display the kinematical distributions, since they are all similar to the left plot of Fig.~\ref{fig:kinematical_distributions}.
The trigger efficiency $\epsilon_{\text{trigger}}$ is independent of the neutrino detectors and is estimated to be about $40\%$ for all the cases.
As for the detector acceptance efficiency $\epsilon_{\text{window}}$, we first show in the right panel of Fig.~\ref{fig:kinematical_distributions} distributions of $\nu_e$ with mass 0.1 eV in the plane $\eta_{\nu_e}$ vs.~$E_{\nu_e}$, produced in $W^-$ decays, after the events pass the trigger requirements on the prompt charged leptons.
We observe that the proportion of events with large $\eta_{\nu_e}$ values is rather small.
For $\nu_e$ from $W^+$ decays, and for $\nu_\mu$'s, the kinematical distributions are also similar, and hence are not presented here.
We list the values of $\epsilon_{\text{window}}$ for each detector in Table~\ref{tab:epsilon_window}.
\begin{table}[t]
\centering
\begin{tabular}{c|cccc}
\hline
\hline
$\epsilon_{\text{window}}$                                                                 & $e^+$                                             & $e^-$                          & $\mu^+$    & $\mu^-$  \\ \hline
FASER$\nu$    & $1.3\times 10^{-6}$     & $6.0\times 10^{-7}$      &$1.7\times 10^{-6}$     & $5.8\times 10^{-7}$                                    \\ 
SND@LHC        & $1.5\times10^{-5}$   & $6.0\times 10^{-6}$        & $2.0\times 10^{-5}$      & $4.6\times 10^{-6}$                                          \\
\hline 
FASER$\nu$2  & $1.3\times 10^{-6}$      & $6.0\times 10^{-7}$        & $1.7\times 10^{-6}$   & $5.8\times 10^{-7}$                                         \\
AdvSND(far)       & $1.5\times 10^{-5}$     &$6.0\times 10^{-6}$   & $2.0\times 10^{-5}$   & $4.6\times 10^{-6}$                                       \\ 
FLArE-10    &  $7.6\times 10^{-6}$   & $5.4\times 10^{-6}$       & $1.3\times 10^{-5}$     & $1.7\times 10^{-6}$                  \\ 
FLArE-100      &$2.7\times 10^{-5}$   & $8.9\times 10^{-6}$        & $3.3\times 10^{-5}$   & $8.1\times 10^{-6}$     \\
\hline
\hline
\end{tabular}
\caption{Table of values of $\epsilon_{\text{window}}$ computed with \texttt{Pythia8} simulation of 10 million $W$-boson production events for each experiment, corresponding to observing same-sign $e^+$, $e^-$, $\mu^+$, and $\mu^-$, respectively. The computation of $\epsilon_{\text{window}}$ is based on the number of events passing the prompt-lepton trigger requirement.}
\label{tab:epsilon_window}
\end{table}

The calculation of the helicity flip was already explained in Sec.~\ref{sec:exp}, and finally, in order to compute $P_{\text{scatt.}}$ according to Eq.~\eqref{eq:Pscatt} we need to obtain the neutrino-nucleus CC scattering cross sections as functions of the neutrino energy $E_\nu$.
This is extracted from Fig.~5 of Ref.~\cite{FASER:2019dxq} for neutrino-tungsten CC scattering cross sections; for neutrino-argon cross sections, we just simply re-scale the neutrino-tungsten scattering values according to the ratio of the tungsten and argon atomic mass numbers as a good approximation.

The final numerical results are summarized in Table~\ref{tab:NSv}.
\begin{table}[t]
\centering
\begin{tabular}{c|cccc}
\hline
\hline
$N_S^\nu$                                                                 & $e^+$                                             & $e^-$      & $\mu^+$    & $\mu^-$  \\ \hline
FASER$\nu$    & $1.1\times 10^{-32}$      & $7.0\times 10^{-32}$     &$3.1\times 10^{-32}$    & $3.4\times 10^{-32}$                                      \\ 
SND@LHC    & $7.5\times10^{-32}$   & $9.0\times 10^{-32}$        & $7.6\times 10^{-32}$     & $8.4\times 10^{-32}$                                          \\
\hline 
FASER$\nu$2        & $1.5 \times 10^{-30}$     & $9.1\times 10^{-31}$    & $4.1\times 10^{-30}$  & $4.4\times 10^{-30}$                                      \\
AdvSND(far)       & $2.6\times 10^{-30}$   &$3.1\times 10^{-30}$   & $2.6\times 10^{-30}$     & $2.9\times 10^{-30}$                                       \\ 
FLArE-10     &  $7.6\times 10^{-31}$   & $3.1\times 10^{-30}$     & $1.6\times 10^{-30}$   & $1.7\times 10^{-30}$          \\ 
FLArE-100      &$1.8\times 10^{-29}$  & $2.0\times 10^{-29}$   & $1.5\times 10^{-29}$   & $1.8\times 10^{-29}$      \\
\hline
\hline
\end{tabular}
\caption{Table of numerical results of $N_S^\nu$ at the neutrino detectors, for light neutrinos' both production and scattering with the SM CC interactions.}
\label{tab:NSv}
\end{table}
One easily sees that the expected signal-event numbers are below $\mathcal{O}(1)$ by about 29 to 32 orders of magnitude.
This means that it is, unfortunately, impossible to observe such LNV signatures with Majorana light neutrinos undergoing purely SM weak-current interactions at these neutrino detectors.

Since the main cause for the extremely small signal-event rates is the strong suppression effect of helicity flip, we proceed to consider a theoretical case where a higher-dimensional operator leads to either the production or the scattering (but not both) of the light neutrino, which is itself lepton-number violating, thus circumventing the helicity flip requirement.
Indeed, physics beyond the Standard Model may induce new interactions mediated by heavy fields which are not directly observable at colliders presently.
However, at relatively low energy scales, such heavy fields are integrated out in the theory and their effects can be encoded in so-called ``Wilson coefficiencts'' of non-renormalizable operators of mass dimension larger than four.
A general framework with such operators including the neutrinos is known as the Standard Model Effective Field Theory (SMEFT) (see Ref.~\cite{Brivio:2017vri} for a review).
In particular, SMEFT higher-dimensional operators which are LNV and include two quarks, one charged lepton, and a neutrino, arise at mass dimension 7~\cite{Babu:2001ex,Lehman:2014jma}.\footnote{Note that such dim-7 operators can also be probed at colliders with processes such as $pp\to l+$ missing energy, cf.~Ref.~\cite{Fridell:2023rtr} for a comprehensive study on LNV with dim-7 operators in the SMEFT.}
Here, we focus on the operator $\epsilon_{ij} (L^i C \gamma_\mu e) (\overline{d}\gamma^\mu u) H^j$ with Wilson coefficient labeled as $1/\Lambda^3$, where $H$ is the Higgs doublet, $\epsilon_{ij}$ is the $SU(2)$ index tensor, $C$ is the Dirac charge conjugation matrix, $\Lambda$ denotes the new-physics scale, leading to the neutrino-nucleus scattering into an electron or muon.
As for the production of the light neutrino, we stick to the leptonic decays of the $W$-boson.
In order to compute the the scattering cross sections of neutrino and nucleons with the dim-7 operator, we apply the MC generator \texttt{MadGraph5\_aMC\_v3.4.1}~\cite{Alwall:2007st,Alwall:2011uj,Alwall:2014hca,Frederix:2018nkq}.
However, \texttt{MadGraph5} does not support multi-fermion operators with at least one Majorana field.
To solve the issue, we implement a UV-complete model of leptoquarks, setting up large leptoquark masses and small decay widths (see also Ref.~\cite{Cottin:2021lzz}).
This would allow us to effectively emulate the dim-7 operator.
Further, for this dim-7 operator, we consider only the first-generation quarks.
The reason is two-fold.
Firstly, the first-generation quarks are the main matter content of nucleons, leading to the largest scattering cross sections. 
Second, such dim-7 operators could induce radiatively neutrino masses, the strength of which is proportional to the masses of the fermion fields~\cite{Kanemura:2010bq,Cepedello:2017eqf}.
Therefore, restricting to the first-generation quarks contains the radiatively generated neutrino masses to a low level obeying the neutrino mass constraints.
For concrete details, we refer the reader to Appendix~\ref{app:LQ}.

Since the production is still through $W$-boson decays, $\sigma_\nu$, $\epsilon_{\text{trigger}}$, as well as $\epsilon_{\text{window}}$ remain unchanged.
However, the helicity flip suppression should not be included: $\epsilon_{\text{h.~f.}}=1$, since the scattering operator itself violates the lepton number.
Finally, $P_{\text{scatt.}}$ is obtained with the newly computed neutrino-nucleus CC scattering cross section $\sigma_{\nu Z}$.
The numerical results of $N_S^\nu$ are given in Table~\ref{tab:NSv-dim7}, for $\Lambda\sim 5$ TeV\,\footnote{If only the electrons are involved, the current bounds on the operator's scale is already close to 50 TeV obtained from neutrinoless double beta decays~\cite{Scholer:2023bnn}. However, when muons are considered, these bounds do not apply. Therefore, we still show these results for a $\Lambda$ of 5 TeV.}.
\begin{table}[t]
\centering
\begin{tabular}{c|cccc}
\hline
\hline
$N_S^\nu$                                                                 & $e^+$                                             & $e^-$                          & $\mu^+$    & $\mu^-$  \\ 
\hline
FASER$\nu$    & $1.5\times 10^{-14}$  & $8.5\times 10^{-15}$  &$7.7\times 10^{-15}$    & $2.0\times 10^{-15}$     \\ 
SND@LHC  & $2.2\times10^{-14}$  & $1.5\times 10^{-14}$      & $3.7\times 10^{-14}$      & $9.3\times 10^{-15}$                                          \\
\hline 
FASER$\nu$2     & $2.0\times 10^{-12}$    & $1.3\times 10^{-12}$ & $1.0\times 10^{-12}$   & $2.4 \times 10^{-13}$                                   \\
AdvSND(far)       & $7.7\times 10^{-13}$     &$5.0\times 10^{-13}$   & $1.2\times 10^{-12}$    & $2.1\times 10^{-13}$                                       \\ 
FLArE-10   &  $6.6\times 10^{-13}$  & $3.6\times 10^{-13}$  & $1.0\times 10^{-12}$  & $4.6\times 10^{-14}$                  \\ 
FLArE-100      &$6.2\times 10^{-12}$    & $2.9\times 10^{-12}$ & $8.5\times 10^{-12}$  & $1.3\times 10^{-12}$      \\
\hline
\hline
\end{tabular}
\caption{Table of numerical results of $N_S^\nu$ for light neutrino produced via an $s$-channel $W$-boson decay and scattering via the dim-7 LNV operator $\epsilon_{ij} (L^i C \gamma_\mu e) (\overline{d}\gamma^\mu u) H^j$ with a coefficient $1/\Lambda^3\sim 1/(5\text{ TeV})^3$, where the neutrino helicity flip is not included.}
\label{tab:NSv-dim7}
\end{table}
We find that even though the strong helicity flip suppression is not present in this case, the estimated signal-event numbers are still more than ten orders of magnitude below $\mathcal{O}(1)$ for a new-physics scale of about 5 TeV.
Since the signal-event numbers are proportional to $\Lambda^{-6}$, even if converting to $\Lambda=1$ TeV, the results would still be several orders of magnitude below 1.

We comment that the existing proposals of using these neutrino detectors to study high-energy neutrinos at colliders only consider primarily neutrinos produced from decays of pions, kaons, and $D$-mesons in an inclusive way.
The initial lepton number in these mesons' decays is hence unknown, not to mention the correlation between the events at the IP and at the neutrino detectors.
Therefore, these proposals in the current shape are unable to determine if the active neutrinos are Majorana or Dirac.
Our proposed search is therefore required to achieve the purpose.

\subsection{Long-lived sterile neutrinos}\label{sec:HNL}

We now study GeV-scale sterile neutrinos produced from $W$-boson decays that mix with active neutrinos.
The SM Lagrangian for the CC and NC interactions is given in Eq.~\eqref{eq:ccnc}.
\begin{eqnarray}\label{eq:ccnc}
{\cal L} &=& \frac{g}{\sqrt{2}}\, 
 V_{\alpha j}\ \bar l_\alpha \gamma^{\mu} P_L N_j W^-_{\mu} +\frac{g}{2 \cos\theta_W}\ \sum_{\alpha, i, j}V^{L}_{\alpha i} V_{\alpha j}^*  
\overline{N_j} \gamma^{\mu} P_L \nu_{i} Z_{\mu} + \text{h.c.},
\end{eqnarray}
where $V^L$ is the PMNS mixing matrix and $\theta_W$ is the Weinberg weak-mixing angle.
For small mixing parameters with the active neutrinos, GeV-scale sterile neutrinos are long-lived and may be observed at the proposed far detectors such as FASER, MATHUSLA, and ANUBIS.
If the sterile neutrino is Majorana, it may decay to a charged lepton of the same sign as that of the mother $W$-boson, leading to LNV signatures.
To observe LNV this way, it is required to achieve event correlation between the prompt and displaced particles.
Moreover, the four-momenta and charge sign of the final-state particles should be measured, where $N$ should decay semi-leptonically into a charged lepton plus hadronic final-state particles.
Since we only consider the sterile neutrino mixed with the electron neutrino, the whole signal process is $pp\to W^\pm \to e^\pm N, N\to e^\pm M^\mp$ or $e^\pm jj$.
The mass threshold determining whether the light sterile neutrino decays into a charged lepton plus a meson or two jets is about 1 GeV~\cite{Bondarenko:2018ptm,DeVries:2020jbs}.

The trigger efficiency $\epsilon_{\text{trigger}}$ is around $40\%$ for the whole range of the considered sterile neutrino mass between about 0.2 GeV and 5 GeV.
Further, for BR$(N\to e^\pm jj/M^\mp)$ as a function of $m_N$, one can find plots in e.g.~Refs.~\cite{Bondarenko:2018ptm,DeVries:2020jbs}; for almost the whole considered mass range, BR$(N\to e^+ jj/M^-)$ is between $20\%$ and $30\%$.
The partial and total decay widths are computed according to formulas given in Refs.~\cite{Bondarenko:2018ptm,DeVries:2020jbs} and the kinematics of the sterile neutrinos are retrieved from \texttt{Pythia8} simulation of 10 million events for each scanned mass value.

\begin{figure}[t]
\centering
 \includegraphics[width=0.8\textwidth]{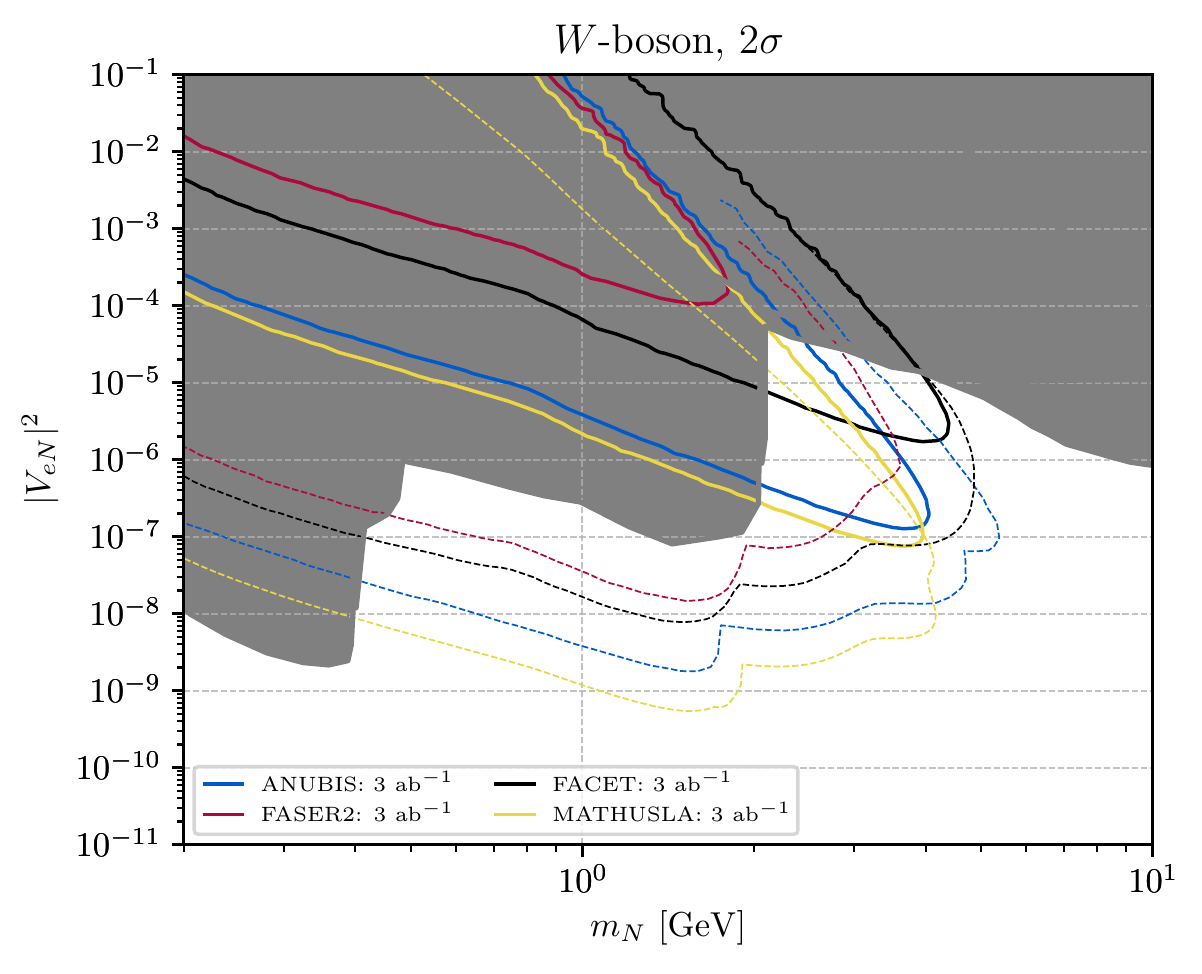}
  \caption{$2\sigma$ exclusion-bound plot for the displaced-vertex search associated with same-sign prompt and displaced leptons, produced from a $W$-boson decay at the IP and from the $N$ decay in the far detectors, respectively.
  The gray area is the currently excluded parameter region, extracted from Refs.~\cite{CHARM:1985nku,Bernardi:1987ek,Baranov:1992vq,DELPHI:1996qcc,CMS:2022fut}.
  The solid lines are the sensitivity reach of the considered experiments with the LNV searches from this study, while the dashed lines correspond to inclusive displaced-vertex searches for HNLs produced from $D$ and $B$-mesons' decays, as well as $W$-boson decays (important for ANUBIS and MATHUSLA only), extracted from Refs.~\cite{DeVries:2020jbs,Beltran:2023nli,Hirsch:2020klk}.
We emphasize that the latter results cannot confirm LNV of the signal events and hence determine the Majorana nature of the HNLs.
  }
  \label{fig:displaced}
\end{figure}
We present the $2\sigma$ ($N_S^N=3$) sensitivity bounds in Fig.~\ref{fig:displaced}.
We plot the sensitivity reach of the considered experiments: FASER2, ANUBIS, MATHUSLA, and FACET, in solid-line style based on this work, while the dashed lines are the projected exclusion bounds with inclusive displaced-vertex searches at the same experiments, for HNLs produced from charm and bottom mesons' decays, as well as $W$-boson decays (important for ANUBIS and MATHUSLA only)~\cite{DeVries:2020jbs,Beltran:2023nli,Hirsch:2020klk}.\footnote{The results of the inclusive displaced-search for HNLs at MATHUSLA were obtained in Ref.~\cite{Hirsch:2020klk} for a slightly different geometrical configuration than the one considered in this work, and therefore the corresponding dashed curve does not completely enclose the solid one in the upper right part of the shown parameter space.}.
The gray area has been excluded by past and existing experiments~\cite{CHARM:1985nku,Bernardi:1987ek,Baranov:1992vq,DELPHI:1996qcc,CMS:2022fut}.
We find that the parameter space that FASER2 is sensitive to has been completely ruled out, and ANUBIS, however, can probe large parts of the unexcluded parameter space.
Although FASER is experimentally capable of performing the search, it does not have sensitivities; this is mainly because of its very small volume and very forward position, and the fact that the sterile neutrinos considered here are produced from $W$-boson which is relatively heavy making the sterile neutrinos travel in a not so forward direction.
Compared to searches for a Majorana sterile neutrino at ATLAS or CMS~\cite{ATLAS:2018dcj,CMS:2018jxx}, our search is sensitive to sterile neutrino mass $\mathcal{O}(1)$ GeV, complementing the sensitive mass reach of these local detector searches which is between 20 GeV and multi-TeV.

In addition, we observe in the plot, as expected, that the exclusion bounds that could be obtained with searches for displaced vertices originating from HNLs in meson decays can cover a larger parameter region for a similar sensitive mass range (see also Refs.~\cite{Helo:2018qej,Curtin:2018mvb,Kling:2018wct,Hirsch:2020klk}).
However, these proposed searches only look for displaced vertices of HNL decays inclusively, without knowledge of the initial lepton number in the meson decays.
Therefore, it is difficult to determine lepton number conservation or violation in the whole process.
As a result, even in the case that signal events of simple displaced-vertex searches are observed, our proposed search can help pin down the Majorana or Dirac nature of the long-lived sterile neutrino.
Concretely speaking, the comparison shown in Fig.~\ref{fig:displaced} clearly highlights the parts of the parameter space where LNV can and cannot be established, respectively, once displaced-vertex signatures are observed at these experiments.

Furthermore, neutrinoless double beta decays ($0\nu\beta\beta$), as the most sensitive test of LNV~\cite{Agostini:2022zub}, are relevant to the study here.
The current lower bounds on the lifetime of $0\nu\beta\beta$ is in the order of magnitude $\mathcal{O}(10^{26}\text{ years})$~\cite{KamLAND-Zen:2022tow,GERDA:2020xhi}.
These bounds can be used for constraining sterile neutrinos; see e.g.~Refs.~\cite{Asaka:2021hkg,deVries:2022nyh,Dekens:2021qch,Bolton:2022tds,Dekens:2020ttz,Dekens:2023iyc}.
In particular, Ref.~\cite{Dekens:2023iyc} studied a scenario closest to ours, where the authors consider a sterile neutrino in the same mass range as ours in a ``3+1'' scenario.
However, they assumed the type-I seesaw relation with the active neutrino mass fixed at 0.05 eV, while we set the mixing angle and the HNL mass as independent parameters.
Their results show that the predicted $0\nu\beta\beta$ lifetime of $^{136}$Xe is $1-2$ orders of magnitude beyond the current leading bound of $2.3\times 10^{26}$ years~\cite{KamLAND-Zen:2022tow}.
Since the computation of the $0\nu\beta\beta$ lifetime of $^{136}$Xe to next-to-next-to-leading order depends on multiple parameters in a highly non-trivial way such as the active-sterile neutrino mixing, sign of the mixing squared, unitarity of the extended PMNS matrix, and the active neutrino mass, a detailed analysis is clearly beyond the scope of this work.
Instead, here we comment only that it is possible that searches for neutrinoless double beta decays can be sensitive to parts of the parameter space overlapping with that which our proposed searches can probe.

Finally, we comment that our results apply almost equally for a sterile neutrino that instead mixes with the muon neutrino dominantly, while for $N$ that mixes only with the tau neutrino, sensitivities should be weakened for reasons of kinematics and $\tau$-reconstruction efficiencies.

\section{Conclusions}\label{sec:conclusions}

Observation of lepton number violation (LNV) would directly point to the Majorana nature of neutrinos.
In this work, we have proposed search strategies for LNV signatures related to Majorana neutrinos, for the first time making use of proposed LHC far detectors that are able to study light neutrinos or displaced decays of long-lived particles (LLPs).
We note that the so-far published phenomenological studies on these LHC far detector have not proposed or studied the search strategies we consider in this work for observing LNV phenomena.

For light neutrinos, we consider production from $W$-boson decays in association with a prompt charged lepton as a trigger, and subsequent neutrino-nucleus charged-current (CC) scattering at neutrino detectors proposed in the forward direction of the ATLAS interaction point (IP).
If the prompt and scattered charged leptons are of the same sign, LNV can be determined, under the conditions that the four-momenta and sign of these leptons as well as the nuclear recoil can be well measured, and the event correlation between the two leptons can be realized.
This is experimentally challenging to achieve, but not impossible for these neutrino detectors.
With the help of Monte-Carlo simulation tool \texttt{Pythia8}, we estimate the signal-event numbers (for same-sign same-flavor charged leptons) at a series of neutrino detectors to be almost negligible, with purely Standard Model (SM) weak-current interactions.

To alleviate the main issue that stems from the helicity flip suppression, we resort to Standard Model Effective Field Theory (SMEFT) considering as an example a dimension-7 LNV operator for inducing the CC scattering, where the production of the active neutrinos is still via the SM $W$-boson and no helicity flip should be included.
The charged-current scattering cross sections are computed with \texttt{MadGraph5} using a UV-complete leptoquark model for emulating the effective operator.
Nevertheless, we find that since the scattering cross section is reduced compared to the purely SM weak-current case, the predicted numbers of signal events are still orders of magnitude below 1 for a new-physics scale $\Lambda$ of 5 TeV.

Despite the negative results of the neutrino-nucleus scattering search, our displaced-decay search at ATLAS/CMS LLP far detectors proves able to probe large new parts of the parameter regions of a GeV-scale long-lived sterile neutrino which mixes with the electron neutrino only.
Similar to the neutrino-nucleus scattering study, we consider the sterile neutrinos produced from leptonic decays of the $W$-bosons in association with a prompt charged lepton which should pass the trigger requirements.
The sterile neutrinos travel a macroscopic distance and potentially decay inside an LLP far detector into a charged lepton plus two jets or a meson, allowing for full reconstruction of the sterile neutrino if the far detector can measure the four-momenta of the final-state particles.
Further, as in the neutrino-nucleus scattering search, it is required to achieve event correlation and determine the lepton charge sign; this imposes strong requirement on the hardware of both the near and far detectors.
We perform MC simulation with \texttt{Pythia8} to determine numerically the expected number of signal events at ANUBIS, FASER2, MATHUSLA, and FACET, even though it is unrealistic to perform the search with the latter two experiments for reasons of either cost or detector location.
We find that while FASER2 can only probe parameter regions that are already excluded, ANUBIS would be sensitive to a large unexcluded parameter space, in search of LNV associated with a GeV-scale sterile neutrino.
In principle, the far detectors proposed in the vicinity of the LHCb experiment could also be used for searching for a similar signature, where now the sterile neutrino should be produced from decays of charm or bottom mesons.
However, since CODEX-b is situated at the small absolute pseudorapidity range and MoEDAL-MAPP would be in the hemisphere of negative pseudorapidity while the LHCb near detector covers the pseudorapidity range of $2-5$, it is difficult for perform a similar search with the LHCb far detectors, since the sterile neutrinos from meson decays are soft and travel in the forward direction.

To summarize the work, we have proposed searches for certain LNV signatures at the LHC far detectors, for which the background is expected to be vanishing.
Such searches are quite challenging in both aspects of hardware and software.
Our numerical computation shows that although it would be difficult for the neutrino detectors to observe an LNV signature associated with a Majorana light neutrino, the LLP far detectors such as ANUBIS may be able to discover an LNV signature arising from a long-lived sterile neutrino of mass $\mathcal{O}(1)$ GeV, if a few events can be observed.
The latter results highly motivate the installation of relevant hardware at LLP far detectors and the performance of such searches.

\section*{Acknowledgements:} 
We thank Felix Kling, Thong T.~Q.~Nguyen, C.~J.~Ouseph, Abner Soffer, Yue-Lin Sming Tsai, and Guanghui Zhou for useful discussions, and thank, in particular, Martin Hirsch and Arsenii Titov, for providing patient guidance on the leptoquark model and for carefully reading the manuscript.
K.W. is supported by the National Natural Science Foundation of China under grant no.~11905162, the Excellent Young Talents Program of the Wuhan University of Technology under grant no.~40122102, and the research program of the Wuhan University of Technology under grant no.~2020IB024.
Y.N.M. is supported by the National Natural Science Foundation of China under grant no.~12205227 and the Fundamental Research Funds for the Central Universities (WUT: 2022IVA052).

\appendix

\section{The dimension-7 operator and leptoquarks} \label{app:LQ} 

Leptoquarks~\cite{Dimopoulos:1979es,Dimopoulos:1979sp,Eichten:1979ah,Angelopoulos:1986uq,Buchmuller:1986iq,Pati:1974yy,Georgi:1974sy,Crivellin:2021ejk} in general couple to both leptons and quarks simultaneously (and hence the name), and can therefore lead to the neutrino-nucleus scattering process via a $t$-channel leptoquark.
Such scattering via a leptoquark has been studied for neutrino detectors at FPF~\cite{Cheung:2023gwm}.
In this appendix we introduce the construction of a UV-complete leptoquark model with two leptoquarks $S_d\sim (\mathbf{3},\mathbf{1})_{-1/3}$ and $S_Q\sim (\mathbf{3},\mathbf{2})_{1/6}$, and the following terms in the Lagrangian:
\begin{eqnarray}
    \mathcal{L}\supset -\Big(  g_{ue} \overline{u_R} e^c_R S_d   + g_{Ld}\overline{d_R} L^T \epsilon S_Q  + \mu S_Q^\dagger S_d H \Big)+\text{h.c.}\ldots,
\label{eq:LQ_Lag}
\end{eqnarray}
where $L$ is the SM left-chiral lepton doublet, $u_R$ are $d_R$ are right-chiral quark singlets, $\epsilon$ is the $SU(2)$ index tensor contracting $L^T$ and $S_Q$, and $H$ is the SM Higgs doublet.
Further, $g_{ue}$ and $g_{Ld}$ are Yukawa-like couplings.
The last term, $\mu S^\dagger_Q S_d H$ with a dimensionful coupling $\mu$, is a bilinear mixing term between the leptoquarks $S_Q$ and $S_d$~\cite{Hirsch:1996qy}, and violates lepton number explicitly.
This mixing term is necessary for our purpose, since we would like to have LNV at the effective operator level.
The mixing matrix between the two leptoquarks can be written as
\begin{eqnarray}
\begin{pmatrix}
\sqrt{1-\frac{v^2 \mu^2}{2(m_d^2-m_Q^2)^2}} & \frac{v\mu}{\sqrt{2}(m_d^2-m_Q^2)} \\
\frac{-v\mu}{\sqrt{2}(m_d^2-m_Q^2)} & \sqrt{1-\frac{v^2 \mu^2}{2(m_d^2-m_Q^2)^2}}
\end{pmatrix}\equiv 
\begin{pmatrix}
    \cos{\theta}&\sin{\theta}\\
    -\sin{\theta} & \cos{\theta}
\end{pmatrix},
\end{eqnarray}
where $v=246$ GeV is the SM Higgs VEV, $m_{d/Q}$ is the Lagrangian-level mass of $S_{d/Q}$, and $\theta$ denotes the leptoquark mixing angle.
This leads to three mass eigenstates of leptoquarks within the model: $\hat{S}_d$, $\hat{S}_{Q^-}$, and $\hat{S}_{Q^+}$, with the following masses:
\begin{eqnarray}
    \hat{m}_d^2&=&m_d^2 + \sin^2{\theta}(m_d^2-m_Q^2),\\
    \hat{m}_{Q^-}^2&=& m_Q^2 - \sin^2{\theta}(m_d^2-m_Q^2),\\
    \hat{m}_{Q^+}^2&=& m^2_Q.
\end{eqnarray}
Note that we have ignored the terms consisting of two Higgs doublets and two leptoquarks (see e.g.~Ref.~\cite{Crivellin:2021ejk}) which can contribute to the leptoquarks' masses.

Currently, the ATLAS and CMS experiments have placed stringent constraints on leptoquarks up to a few TeV~\cite{CMS:2018txo,CMS:2018ncu,CMS:2018lab,ATLAS:2020dsk,CMS:2020wzx,ATLAS:2022wcu}.
Setting both the leptoquark masses $m_d$ and $m_Q$ to be about 10 TeV with $g_{ue}=g_{Ld}=1$, allows us to integrate out the heavy leptoquark fields, reaching the dim-7 LNV operator $\epsilon_{ij} (L^i C \gamma_\mu e) (\overline{d}\gamma^\mu u) H^j$\,\footnote{It is important to have heavy leptoquarks so as to avoid their $s$-channel on-shell production at the LHC which would make integrating out the heavy leptoquarks invalid for our purpose.}.
The corresponding Wilson coefficient is
\begin{eqnarray}
\frac{1}{\Lambda^3}=\frac{\mu g_{ue} g_{Ld}}{2\, m_Q^2 m_d^2},
\label{eq:WC_matching}
\end{eqnarray}
where the factor 2 in the denominator arises from Fierz identities~\cite{Liao:2016hru}.
Other dim-7 LNV operators such as $ \epsilon_{ij}\epsilon_{mn} (\overline{d}L^i)(Q^j C  L^m) H^n$ could be induced, if a further term $g_{QL} \Bar{Q}\epsilon L^c S_d $ is also present in the theory, where $Q$ is the SM left-chiral quark doublet.

We implement the leptoquark model~\eqref{eq:LQ_Lag} in UFO format~\cite{Degrande:2011ua} with FeynRules~\cite{Alloul:2013bka}, including setting the active neutrinos to be Majorana, and generate the signal events with \texttt{MadGraph5} for computing the neutrino-proton/neutron scattering cross sections: $p\, \nu_{e/\mu} \to j e^+/\mu^+$ and $p\,\nu_{e/\mu}\to j e^-/\mu^-$, fixing the neutrino beam polarization to be $100\%$ left-handed and right-handed, respectively, and setting the incoming nucleon beam at rest.
We turn on only couplings with the first-generation quarks and leptons of either the first or the second generation (but not both), and set them to 1.
The input Lagrangian-level leptoquark masses $m_d$ and $m_Q$ are fixed to be 10000 GeV and 9999 GeV, with the mixing angle $\sin{\theta}$ set to be $\frac{1}{\sqrt{2}}$. 
The corresponding Wilson coefficient can be calculated with Eq.~\eqref{eq:WC_matching} to be $1/\Lambda^3 = 1/(62.66\text{ TeV})^3$.
We thus obtain the neutrino-proton scattering cross sections with \texttt{MadGraph5} with all parton-level kinematic cuts removed, as a function of the incoming neutrino energy.
As for calculating neutrino-neutron scattering cross sections, we generate the same processes, but set the proton `\textit{p}' to be a pure neutron with the correct mass 939.57 MeV.
Note that scattering cross sections peak, for $E_\nu\sim$ TeV, at the momentum transferred $Q^2=2\,x\,y\,E_\nu m_{p/n}$~\cite{FASER:2019dxq}, with $x=0.1$ being the fraction of the proton’s or neutron's momentum carried by the quark in the initial state, $y=0.5$ being the fraction of the neutrino momentum transferred to the hadronic system, $E_\nu$ being the neutrino energy, and $m_{p/n}$ being the mass of a proton or neutron.
Correspondingly, in \texttt{run\_card.dat} of \texttt{MadGraph5}, we tune the renormalization scale and the factorization scale of the proton/neutron target to be $\sqrt{Q^2}$.
We then scan over the neutrino energy range $[10\text{ GeV}, 10\text{ TeV}]$ and attain a list of scattering cross sections with the dim-7 operator.
\begin{figure}[t]
\centering
 \includegraphics[width=0.5\textwidth]{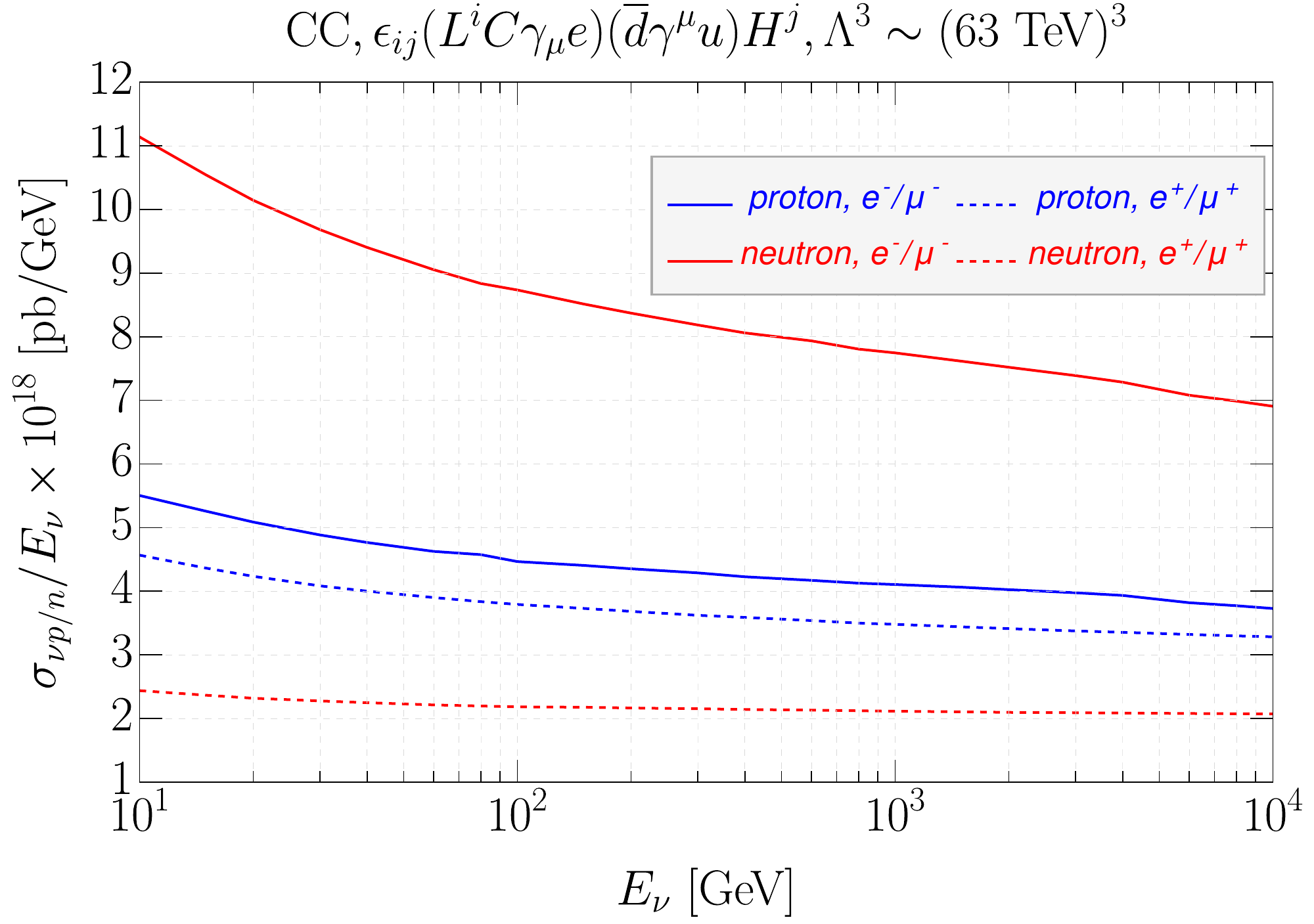}
  \caption{Neutrino-nucleon charged-current scattering cross sections divided by $E_\nu$, as functions of $E_\nu$, via the dim-7 operator $\epsilon_{ij} (L^i C \gamma_\mu e) (\overline{d}\gamma^\mu u) H^j$, for a Wilson coefficient of $1/\Lambda^3 \sim 1/(63\text{ TeV})^3$.
  Only the first-generation quarks enter the interaction and the new-physics scale $\Lambda$ is approximately 63 TeV.
  }
  \label{fig:vN_XS}
\end{figure}
In Fig.~\ref{fig:vN_XS}, we show a plot for neutrino-proton and neutrino-neutron CC scattering cross section via the dim-7 LNV operator for $\Lambda^3\sim (63\text{ TeV})^3$.
As mentioned in Sec.~\ref{sec:v}, $N_S^\nu$ is proportional to $1/\Lambda^6$, and it is hence straightforward to convert these scattering cross sections to those for $\Lambda=5$ TeV corresponding to the signal-event numbers given in Table~\ref{tab:NSv-dim7}.

At the end, to compute the neutrino-nucleus scattering cross section $\sigma_{\nu Z}$, we take the sum of neutrino-proton and neutrino-neutron scattering cross sections weighted by the numbers of protons and neutrons in a (tungsten or argon) nucleus.

\bibliographystyle{JHEP}
\bibliography{main}

\end{document}